\documentstyle[emulateapj,apjfonts,epsf]{article}

\slugcomment{Accepted for publication by The Astrophysical Journal,
April 18, 2000}

\def\dddotnu{\ifmmode\stackrel{_{...}}{\nu}\else$\stackrel{_{...}}{\nu}$\fi}

\begin{document}

\title{Discovery of Two High-Magnetic-Field Radio Pulsars}

\author{F.~Camilo,\altaffilmark{1,2}
  V.~M.~Kaspi,\altaffilmark{3,4,8}
  A.~G.~Lyne,\altaffilmark{2}
  R.~N.~Manchester,\altaffilmark{5}
  J.~F.~Bell,\altaffilmark{5}
  N.~D'Amico,\altaffilmark{6,7}
  N.~P.~F.~McKay,\altaffilmark{2} and
  F.~Crawford\altaffilmark{3}}
\medskip
\altaffiltext{1}{Columbia Astrophysics Laboratory, Columbia
  University, 550 West 120th Street, New York, NY~10027.  E-mail:
  fernando@astro.columbia.edu}
\altaffiltext{2}{University of Manchester, Jodrell Bank Observatory,
  Macclesfield, Cheshire, SK11~9DL, UK}
\altaffiltext{3}{Department of Physics and Center for Space
  Research, Massachusetts Institute of Technology, Cambridge, MA~02139}
\altaffiltext{4}{Department of Physics, Rutherford Physics Building,
  McGill University, 3600 University Street, Montreal, Quebec, H3A~2T8,
  Canada}
\altaffiltext{5}{Australia Telescope National Facility, CSIRO,
  P.O.~Box~76, Epping, NSW~1710, Australia}
\altaffiltext{6}{Osservatorio Astronomico di Bologna, via
  Ranzani 1, 40127~Bologna, Italy}
\altaffiltext{7}{Istituto di Radioastronomia del CNR, via Gobetti
  101, 40129~Bologna, Italy}
\altaffiltext{8}{Alfred P. Sloan Research Fellow}

\begin{abstract}
We report the discovery of two young isolated radio pulsars with very
high inferred magnetic fields.  PSR~J1119$-$6127 has period
$P=0.407$\,s, and the largest period derivative known among radio
pulsars, $\dot P=4.0\times10^{-12}$.  Under standard assumptions these
parameters imply a characteristic spin-down age of only $\tau_{\rm c} =
1.6$\,kyr and a surface dipole magnetic field strength of
$B=4.1\times10^{13}$\,G.  We have measured a stationary
period-second-derivative for this pulsar, resulting in a braking index
of $n=2.91\pm0.05$.  We have also observed a glitch in the rotation of
the pulsar, with fractional period change $\Delta P/P = -
4.4\times10^{-9}$.  Archival radio imaging data suggest the presence of
a previously uncataloged supernova remnant centered on the pulsar.  The
second pulsar, PSR~J1814$-$1744, has $P=3.975$\,s and $\dot
P=7.4\times10^{-13}$.  These parameters imply $\tau_{\rm c} = 85$\,kyr,
and $B=5.5\times10^{13}$\,G, the largest of any known radio pulsar.

Both PSR~J1119$-$6127 and PSR~J1814$-$1744 show apparently normal radio
emission in a regime of magnetic field strength where some models
predict that no emission should occur.  Also, PSR~J1814$-$1744 has spin
parameters similar to the anomalous X-ray pulsar (AXP) 1E~2259+586, but
shows no discernible X-ray emission.  If AXPs are isolated, high
magnetic field neutron stars (``magnetars''), these results suggest
that their unusual attributes are unlikely to be merely a consequence
of their very high inferred magnetic fields.

\end{abstract}


\section{Introduction}\label{sec:intro}

The pulsar in the Crab nebula (PSR~B0531+21), with period $P = 33$\,ms,
was born in a type~II supernova observed in 1054~AD, supporting the
view that at least some core-collapse supernovae (SNe) form pulsars.
Based largely on studies of the Crab and a few other young objects, a
picture has emerged where pulsars are born spinning rapidly (with
initial period $P_0 \approx 20$\,ms in the case of the Crab), and spin
down due to their large magnetic moments according to $\dot{\nu}
\propto - \nu^n$.  In this spin-down law $\nu = 1/P$ is the pulsar
rotation frequency, $\dot \nu$ is its derivative, and $n = \nu \ddot
\nu/(\dot \nu)^2$ is the ``braking index.''  Integration of the
spin-down law with constant magnetic moment gives the age of the
pulsar,
\begin{equation}
\label{eq:tau}
\tau = \frac{P}{(n-1) \dot{P}} \left[ 1 - \left(\frac{P_0}{P}
\right)^{n-1} \right].
\end{equation}
Braking indices have been measured for only four pulsars, namely
PSRs~B0531+21, B0540$-$69, B0833$-$45, and B1509$-$58, with values for
$n$ of $2.51\pm0.01$, $2.2\pm0.1$, $1.4\pm0.2$, and $2.837\pm0.001$
respectively (\cite{lps93}; \cite{dnb99}; \cite{lpgc96};
\cite{kms+94}).  In other cases, an oblique rotating vacuum dipole
model is typically assumed, for which $n=3$ (\cite{mt77}), and if $P_0
\ll P$, equation~(\ref{eq:tau}) reduces to $\tau = P/(2 \dot P) \equiv
\tau_{\rm c}$, the characteristic age of a pulsar.  With a neutron star
radius of $10^6$\,cm and moment of inertia of $10^{45}$\,g\,cm$^2$, the
surface magnetic field strength is
\begin{equation}
\label{eq:B}
B = 3.2 \times 10^{19} \sqrt{P \dot{P}} \;\; {\rm G}.
\end{equation}
The luminosity generated in the braking of the pulsar rotation, $\dot E
= 4\pi^2 I \nu \dot \nu$, is emitted in the form of magnetic dipole
radiation and a relativistic particle wind.  The vast majority of this
luminosity may be deposited in the ambient environment, powering a
plerionic supernova remnant (SNR) such as the Crab synchrotron nebula,
while a very small portion may be observed as pulsed electromagnetic
radiation.

Despite the above, many questions remain regarding the outcome of
type~II SNe and the manifestation of young neutron stars.  Although
Galactic SNe and pulsar formation rates are both notoriously difficult
to estimate (see, e.g., \cite{vt91}; \cite{tls94}; \cite{wol98}, and
\cite{no90}; \cite{lbdh93}; \cite{lml+98}), it is quite plausible that
type~II SNe occur significantly more often than radio pulsars of the
kind already known are born (see \cite{vt91}; \cite{wol98}).  If this
is the case, perhaps some young neutron stars are being ``missed.''  A
possible example is SNR 3C58, the likely outcome of a type~II SN
observed about 820\,yr ago, with no detectable pulsar.  Studying the
energetics and morphology of the remnant, Helfand, Becker, \& White
(1995)\nocite{hbw95} make a compelling case for the presence of an
unseen pulsar with higher magnetic field than any previously known.
Having a short period like the Crab at birth, such a pulsar would have
spun down rapidly to a present long period.  Maybe yet other pulsars
are born spinning slowly and never generate the large $\dot E$ required
to power an easily detectable nebula.  In addition, some neutron stars
may never manifest themselves as radio pulsars at all.  It has been
suggested that there exists a class of isolated rotating neutron stars
with ultra-strong magnetic fields, the so-called ``magnetars''
(\cite{dt92a}).  The observational properties of radio pulsars and
magnetar candidates are very different.  Radio pulsars rarely exhibit
X-ray pulsations, and when they do, their X-ray power is small compared
to their $\dot E$.  By contrast, magnetars emit pulsed X-rays with
luminosities far in excess of their spin-down power (\cite{vg97};
\cite{ksh+99}) but remain undetected at radio wavelengths.  The
dichotomy is thought to result from the much larger magnetic fields in
magnetars (\cite{td93a}; \cite{hh97}).

In this paper we report the discovery of two isolated radio pulsars
with some properties that are unusual and interesting in the context of
the above questions.

\section{Observations and Results}\label{sec:obs}

The radio pulsars J1119$-$6127 and J1814$-$1744 were discovered on 1997
August 24 and 23, respectively, in a survey of the Galactic plane using
the 64-m Parkes radio telescope in Australia.  This survey
(\cite{lcm+00}; \cite{clm+00}) makes use of the fast rate of sky
coverage afforded by a multibeam receiver to increase greatly the
integration time, and consequently the sensitivity, relative to
previous surveys.  The Parkes survey uses 13 beams at a central sky
frequency of 1374\,MHz with an equivalent system noise of $S_0 =
35$\,Jy at high Galactic latitude.  For each beam, the sum of two
orthogonal linear polarization channels, each 288\,MHz wide, is
recorded for 35 minutes per grid position, providing sensitivity to all
pulsars with flux densities in excess of $\sim0.15$\,mJy for
$P\ga0.1$\,s.

Follow-on regular timing observations have been carried out at Parkes
since 1998 February for newly discovered pulsars with declination south
of $-35\arcdeg$, while the remainder are observed with the 76-m
telescope at Jodrell Bank Observatory, England.  The system used for
timing observations at Parkes is identical to that used in the survey,
although we record signals from the central beam only: the
down-converted radio-frequency noise is passed through a
$2\times96\times3$-MHz filter bank spectrometer, after which the
signals are square-law detected, orthogonal polarizations are summed,
and the 96 resulting voltages are high-pass filtered before being 1-bit
digitized every 250\,$\mu$s and written to magnetic tape for subsequent
analysis.  We also record the start time of each observation,
synchronized with the observatory time standard and traceable to UTC.
PSR~J1119$-$6127, in whose direction $S_0=40$\,Jy, was observed in this
manner on 63 days over a period of two years, for approximately 10
minutes each day.  At Jodrell Bank, with $S_0=50$\,Jy in the direction
of PSR~J1814$-$1744, the observing setup used a $2\times32\times3$-MHz
filter bank until 1999 July, and a $2\times64\times1$-MHz filter bank
since then, to observe a band centered in the range 1376 to 1396\,MHz,
depending on the radio-frequency interference environment.  Signals
from individual frequency channels are delayed by an amount
proportional to the dispersion measure (DM) of the pulsar, to account
for dispersion caused by propagation through the interstellar medium,
and are folded synchronously with the predicted rotation period,
generating one pulse profile for each sub-integration lasting 3
minutes.  We have observed PSR~J1814$-$1744 in this manner on 37 days
over a two-year interval, for 18 minutes each day.

We have analyzed the timing data in standard fashion.  Briefly,
topocentric pulse times-of-arrival (TOAs) were measured by
cross-correlating daily-averaged pulse profiles with a high
signal-to-noise-ratio template (see Fig.~\ref{fig:profs}).  Celestial
coordinates and spin parameters were then determined using the {\sc
tempo} software package\footnote{http://pulsar.princeton.edu/tempo.}
and the JPL~DE200 planetary ephemeris (\cite{sta90}).  {\sc tempo}
first converts the TOAs to the barycenter of the solar system, and
refines the initial estimated parameters in a fitting procedure that
minimizes timing residuals (difference between observed and predicted
TOAs) with respect to the model parameters.

\medskip
\epsfxsize=8truecm
\epsfbox{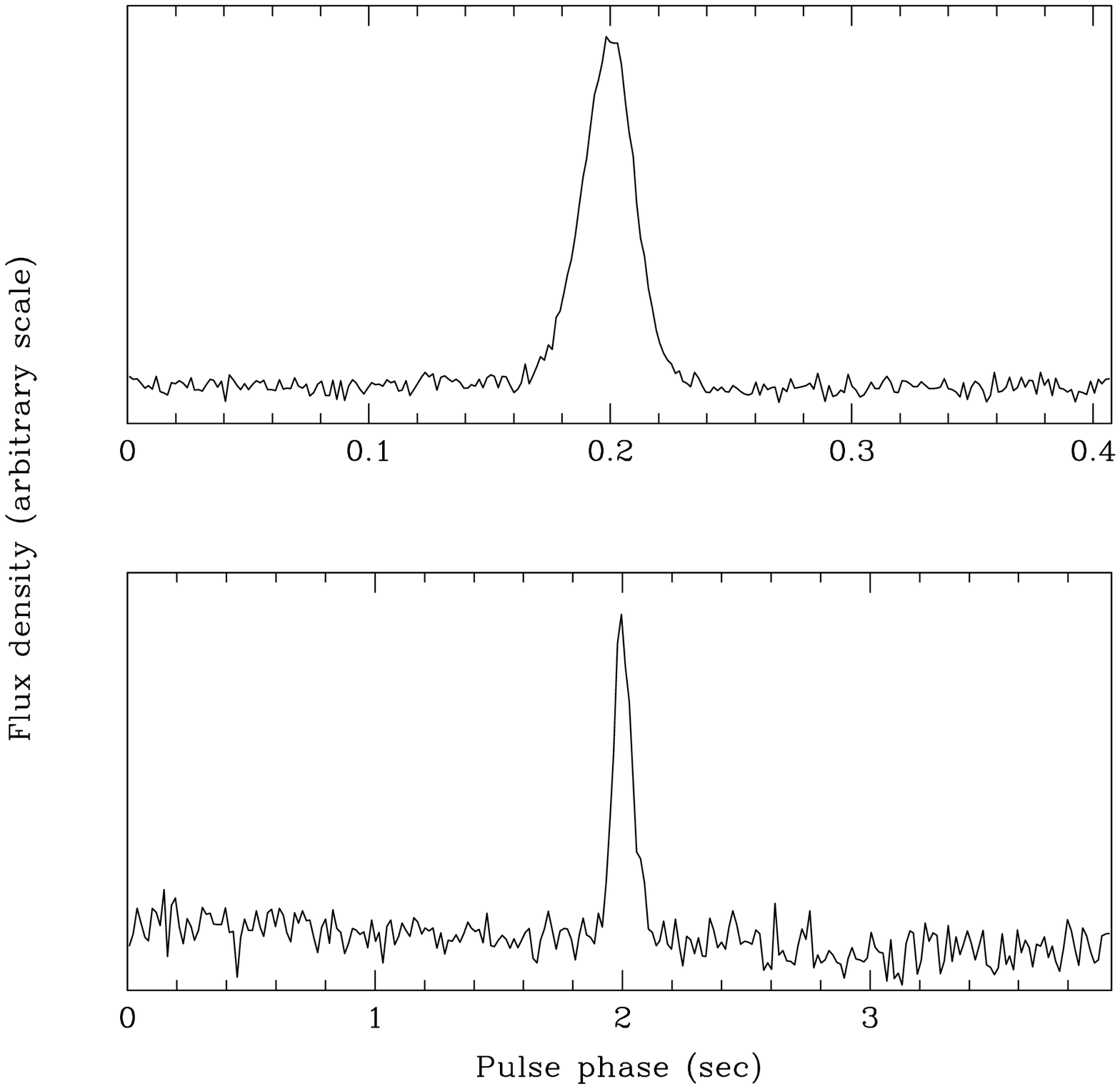}
\figcaption[profs.eps]{\label{fig:profs} 
Integrated pulse profiles for PSRs~J1119$-$6127 (top) and J1814$-$1744
(bottom) at a frequency of 1374\,MHz. }
\bigskip

Underlying the timing model is the assumption that the rotational phase
of the neutron star is described by
\begin{equation}
\label{eq:phase}
\phi(T) = \nu T + \frac{1}{2}\dot\nu T^2 + \frac{1}{6}\ddot\nu T^3 + \ldots,
\end{equation}
where $T$ denotes pulsar proper time.  In equation~(\ref{eq:phase}),
the interpretation of $\nu$ and its derivatives as representing only
the stationary spin parameters of a rotating magnetic dipole does not
hold strictly for pulsars displaying rotational irregularities, and
parameter estimation in such circumstances must be performed with extra
care.

PSR~J1119$-$6127 was observed on 1998 October 30 and 31 with the
Australia Telescope Compact Array (ATCA), with the interferometer in
its ``6D'' configuration (\cite{atca99}).  The observations were done
in pulsar gating mode simultaneously at center frequencies of 1384 and
2496\,MHz, with 128\,MHz of bandwidth in each of two linear
polarizations at each frequency.  The radio sources 1934$-$638 and
1036$-$697 were used as flux density and phase calibrators,
respectively.  The data were processed using the {\sc miriad}
package\footnote{http://www.atnf.csiro.au/computing/software/miriad.},
during which on- and off-pulse maps were generated.  The data set at
1384\,MHz, at which frequency the pulsar is brightest, was used to
obtain the position of the pulsar, and the 2496\,MHz observation was
used to determine the flux density, both listed in
Table~\ref{tab:parms}.

Because PSR~J1119$-$6127 has the largest period derivative ($\dot P$)
of any known radio pulsar, and $\dot P$ correlates highly with ``timing
noise'' (\cite{antt94}) which can bias the celestial coordinates
determined with timing data, we use the position obtained from
interferometric observations in the timing solution.  We then fit for
$\nu$, $\dot \nu$, and $\ddot \nu$.  Figures~\ref{fig:res}{\em a\/} and
{\em b\/} indicate a small glitch in rotation occurred on about MJD
51398; glitch parameters are given in Table~\ref{tab:parms}.
Figure~\ref{fig:res}{\em b\/} also suggests that another glitch of
similar magnitude may have occurred sometime during MJD 50850--50940,
but we cannot be sure.  The rotational parameters best describing the
behavior of the pulsar are listed in Table~\ref{tab:parms}, with the
corresponding timing residuals shown in Figure~\ref{fig:res}{\em c\/}.

\medskip
\epsfxsize=8truecm
\epsfbox{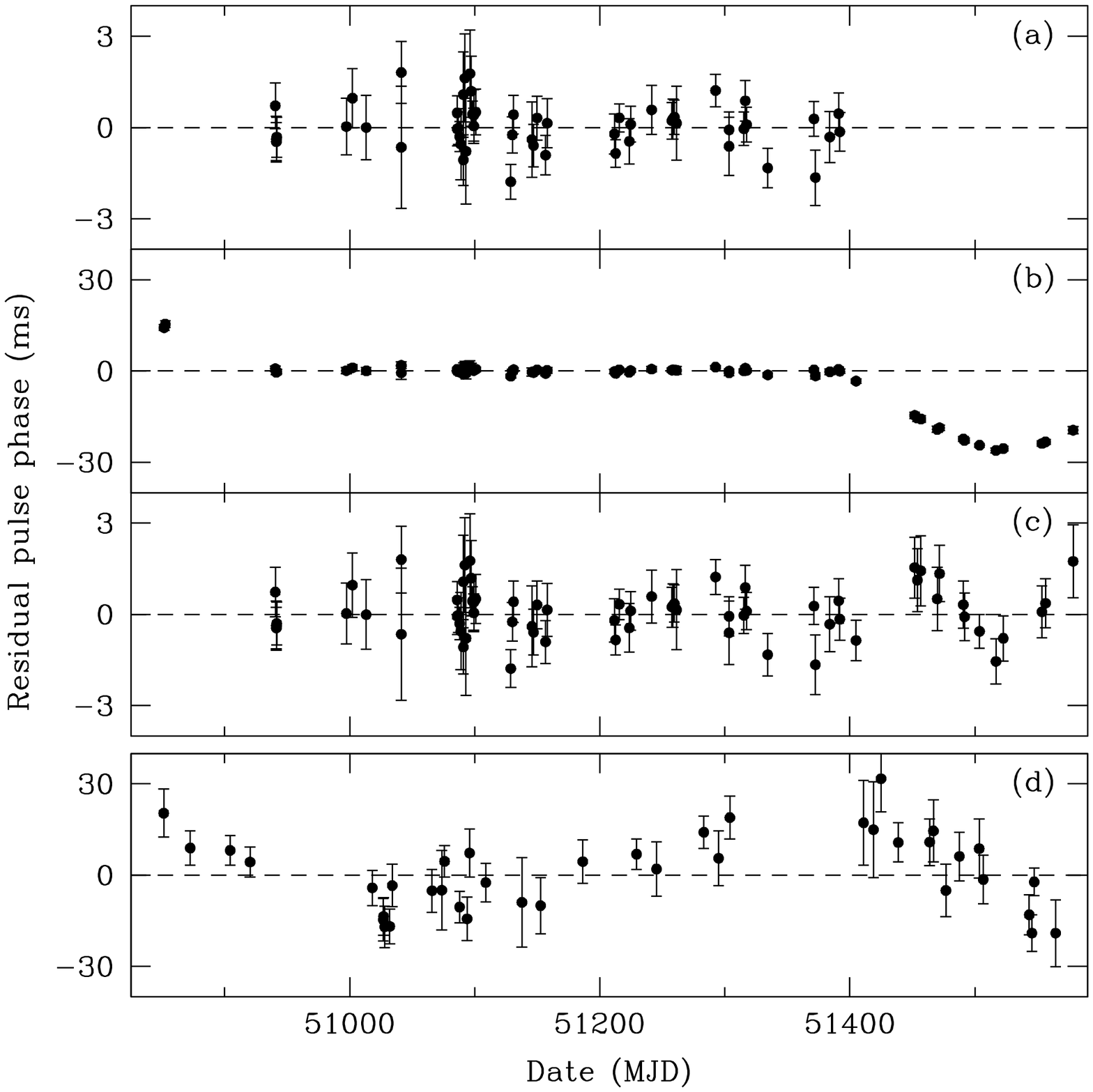}
\figcaption[res.eps]{\label{fig:res} 
(a) Timing residuals for PSR~J1119$-$6127 using data collected over
1998 May--1999 August after fitting for $\nu$, $\dot \nu$, and $\ddot
\nu$.  (b) Residuals for entire data set using model obtained in (a),
showing a glitch at about MJD 51398. (c) Residuals using data collected
over 1998 May--2000 February after fitting for $\nu$, $\dot \nu$,
$\ddot \nu$, and glitch parameters.  (d) Post-fit residuals for
PSR~J1814$-$1744. }
\bigskip

\begin{deluxetable}{lll}
%\tablewidth{0pt}
\scriptsize
\tablecaption{\label{tab:parms}Measured and derived parameters for
pulsars J1119$-$6127 and J1814$-$1744}
\tablecolumns{3}
\tablehead{
\colhead{}                    &
\colhead{PSR~J1119$-$6127}    &
\colhead{PSR~J1814$-$1744} \nl}
\startdata
Right ascension, $\alpha$ (J2000)\dotfill & 11 19 14.30(2)\tablenotemark{a}
                                          & 18 14 42.94(10)                  \nl
Declination, $\delta$ (J2000)\dotfill & $-61$ 27 49.5(2)\tablenotemark{a}
                                      & $-17$ 44 25(19)                      \nl
Rotation frequency, $\nu$ (s$^{-1}$)\dotfill & 2.4531601130(1)
                                             & 0.2515197413(3)               \nl
Frequency derivative, $\dot \nu$ (s$^{-2}$)\dotfill 
                            & $-2.4207996(8)\times10^{-11}$ 
                            & $-4.7002(4)\times10^{-14}$                     \nl
Second frequency derivative, $\ddot \nu$ (s$^{-3}$)\dotfill 
                            & $6.94(2)\times10^{-22}$
                            & $3.0(7)\times10^{-24}$~\tablenotemark{b}       \nl
Third frequency derivative, $\dddotnu$ (s$^{-4}$)\dotfill
                                             & $<10^{-30}$  & \nodata        \nl
Epoch (MJD)\dotfill                          & 51173.0 & 51200.0             \nl
Frequency step at glitch, $\Delta \nu$ (s$^{-1}$)\dotfill
                                        & $1.08(10)\times10^{-8}$ & \nodata  \nl
Change in $\dot \nu$ at glitch, $\Delta \dot \nu$ (s$^{-2}$)\dotfill
                                        & $-9.5(13)\times10^{-16}$ & \nodata \nl
Epoch of glitch (MJD)\dotfill                & 51398(4)     & \nodata        \nl
R.M.S. residual (ms) (white/red)\dotfill     & 0.7/\nodata  & 6.3/11.2       \nl
Dispersion measure, DM (cm$^{-3}$\,pc)\dotfill & 707(2)     & 834(20)        \nl
Flux density at 1374\,MHz, $S$ (mJy)\dotfill & 0.9(1)       & 0.8(1)         \nl
Flux density at 2496\,MHz (mJy)\dotfill &0.44(5)\tablenotemark{a}&\nodata\nl
\tableline
Spin period, $P$ (s)\dotfill        & 0.40763747736(2) & 3.975831061(5)      \nl
Period derivative, $\dot P$\dotfill & $4.022602(2)\times 10^{-12}$
                                    & $7.4297(6)\times 10^{-13}$             \nl
Surface magnetic field, $B$ (Gauss)\dotfill            & $4.1\times 10^{13}$ 
                                                       & $5.5\times 10^{13}$ \nl
Characteristic age, $\tau_{\rm c}$ (kyr)\dotfill       & 1.6       & 85      \nl
Spin-down luminosity, $\dot E$ (erg\,s$^{-1}$)\dotfill & $2.3\times10^{36}$
                                                       & $4.7\times10^{32}$  \nl
Braking index, $n$\dotfill                             & 2.91(1)   & \nodata \nl
Distance, $d$ (kpc)\dotfill                            & 2.4--8    & $10(2)$ \nl
Radio luminosity, $S d^2$ (mJy\,kpc$^2$)\dotfill       & $\sim 25$ &$\sim 80$\nl
Galactic longitude, $l$ ($\deg$)\dotfill               & 292.15    & 13.02   \nl
Galactic latitude, $b$ ($\deg$)\dotfill                & $-0.54$   & $-0.21$ \nl
\enddata

\tablecomments{Units of right ascension are hours, minutes, and
seconds, and units of declination are degrees, arcminutes, and
arcseconds.  Figures in parentheses represent 1\,$\sigma$ uncertainties
in least-significant digits quoted.}

\tablenotetext{a}{Obtained from interferometric observations (see
\S~\ref{sec:obs}).  Celestial coordinates obtained from a fit to data
collected over MJD 50940--51392 are $\alpha = 11^{\rm h}19^{\rm
m}14\fs24(5)$, $\delta = -61\arcdeg27'49\farcs8(5)$.}

\tablenotetext{b}{This parameter is not stationary (see
\S~\ref{sec:obs} for details).}

\end{deluxetable}

In Figure~\ref{fig:res}{\em c\/} the residuals following the glitch
appear cubic in shape, with amplitude much reduced by comparison with
the parabolic residuals in Figure~\ref{fig:res}{\em b\/}.  This
suggests that the glitch parameters in Table~\ref{tab:parms} do not
completely describe the behavior of the pulsar following the event.
Further data are required to determine whether the post-glitch spin
parameters are relaxing with exponential decay time-scales of order
several months, as seen in the Vela pulsar, or whether the change in
$\dot \nu$ at the glitch (or at least much of it) is permanent, as
observed in the Crab pulsar (see \cite{lsg00}).

The uncertainty in the braking index determined from the spin
parameters, $n=2.91\pm0.01$, reflects only the random phase noise
resulting from uncertainty in the TOAs.  We now consider the effect
upon this measurement of the possible presence of timing noise and the
known occurrence of glitch(es).  For most pulsars the stationary value
of $\ddot \nu$ in equation~(\ref{eq:phase}) is too small to be
measured, and in timing fits where only $\nu$ and $\dot \nu$ are
determined, excess noise manifests itself as a quasi-cubic trend in the
residuals.  Arzoumanian et al.~(1994)\nocite{antt94} use the parameter
\begin{equation}
\label{eq:delta}
\Delta (t) = \log \left( \frac{1}{6\nu} |\ddot \nu| t^3 \right)
\end{equation}
to estimate the cumulative phase contribution over time $t$ due to
timing noise and find that, for $t=10^8$\,s (arbitrary, but similar to
the time span of their observations), most pulsars, despite a large
scatter in the data, follow the relationship
\begin{equation}
\label{eq:delta8}
\Delta_8 = 6.6 +0.6 \log \dot P.
\end{equation}
Using equations~(\ref{eq:delta})--(\ref{eq:delta8}) for
PSR~J1119$-$6127, we estimate that the measured value of $\ddot \nu$
may be contaminated by as much as $8\times10^{-24}\,\mbox{s}^{-3}$, or
four times the formal uncertainty given in Table~\ref{tab:parms}.  In
fact we have not measured any timing noise for this pulsar, as
indicated by the apparently ``white'' residuals (Fig.~\ref{fig:res}{\em
a\/}) and by the upper limit on, rather than measurement of, $\dddotnu$
obtained in an additional fit to the data represented in
Figure~\ref{fig:res}{\em a\/} (see Table~\ref{tab:parms}).  However
this is not too surprising, given that our inter-glitch data span only
1.2\,yr; with a longer time span between glitches, timing noise, if it
is present, may be measurable.  In summary, we believe an accurate
measurement of the braking index between glitches, reflecting the
steady spin-down physics of the neutron star, is
$n=2.91\pm0.01\pm0.04$.

If the change in $\dot \nu$ at the glitch, $\Delta \dot \nu$
(Table~\ref{tab:parms}), is permanent, it contributes a component to
$\ddot \nu$ beyond that measured between glitches.  This contribution
is approximately $\Delta \dot \nu/\Delta T$, where $\Delta T$ is the
time interval between glitches.  By assuming that all of the measured
$\Delta \dot \nu$ is permanent, and that a first glitch did occur at
$\mbox{MJD} \sim 50900$ (see Fig.~\ref{fig:res}{\em b\/}), we obtain
$\Delta \ddot \nu = -2\times10^{-23}\,\mbox{s}^{-3}$, about 10 times
the formal uncertainty in $\ddot \nu$ and implying a correction to $n$
of $-0.1$.  Future measurements will settle this question, but for the
purposes of calculating pulsar age, the correct value of braking index
may be as low as $\approx 2.8$.  A similar effect is seen in the Crab
pulsar, where permanent changes in spin-down rate occurring at glitches
contribute a correction of $-0.05$ to the value of $n=2.5$ measured
between glitches (\cite{nic93a}).

The quasi-cubic trend in the residuals of PSR~J1814$-$1744 shown in
Figure~\ref{fig:res}{\em d\/} suggests the presence of timing noise in
this pulsar.  In these circumstances, we determined its position by
``whitening'' the residuals with a fit of the data to a model
incorporating celestial coordinates, $\nu$, $\dot \nu$, and $\ddot
\nu$.  We took the resulting coordinates, with uncertainties, as our
best unbiased estimate of position, and kept them fixed in subsequent
fits to the timing data.  We then performed one fit for $\nu$ and $\dot
\nu$, resulting in the best values for these stationary parameters
averaged over the data span, with residuals displayed in
Figure~\ref{fig:res}{\em d\/} and showing some red-noise.  Finally we
performed one extra fit with the additional free parameter $\ddot \nu$,
which is not stationary.  Values obtained for all these parameters are
listed in Table~\ref{tab:parms}.  We note that the value of $\ddot
\nu$, an estimate of the amount of timing noise present in
PSR~J1814$-$1744, is approximately at the level expected from
equations~(\ref{eq:delta})--(\ref{eq:delta8}).  Note that the
uncertainty in declination is particularly large because this pulsar is
located at low ecliptic latitude.

The DMs quoted in Table~\ref{tab:parms} were obtained by folding raw
data at the known pulse period for each of four frequency sub-bands,
created by the addition of data from 24 adjacent frequency channels,
and fitting a time-delay between the sub-bands.  This was done for
timing data in the case of PSR~J1119$-$6127 and discovery data for
PSR~J1814$-$1744.  Dispersion measures, together with a model for the
Galactic electron density distribution (\cite{tc93}), are used to
estimate distances to pulsars.  For PSR~J1119$-$6127 the implied
distance is $d>30$\,kpc.  However, the model does not account for most
individual H{\sc ii} regions, and, because the pulsar lies in the
direction of the Carina spiral arm, with the likelihood of nearby
ionizing population~I stars, this distance is assuredly a gross
overestimate.  We believe it is likely that the pulsar is located
between the two line-of-sight crossings of the Carina arm, between
$d=2.4$ and 8\,kpc.  The distance estimate quoted in
Table~\ref{tab:parms} for PSR~J1814$-$1744 is that obtained from the
model of Taylor \& Cordes (1993)\nocite{tc93}.

The flux densities at 1374\,MHz listed in Table~\ref{tab:parms} were
determined by converting the average observed signal strength of the
pulsars to a scale calibrated using published flux densities at
1400\,MHz for a group of high-DM pulsars, taking into account the
variation in sky background temperature.

\section{Discussion}\label{sec:disc}

\subsection{PSR~J1119$-$6127}\label{sec:p1119}

PSR~J1119$-$6127 has the largest period derivative known among radio
pulsars.  Partly for this reason it was relatively straightforward to
measure a stationary $\ddot \nu$ with a phase-connected timing solution
(i.e., through absolute pulse numbering), only the third pulsar for
which this has been possible.  The resulting value of braking index is
$n=2.91\pm0.05$, including possible contamination by timing noise (see
\S~\ref{sec:obs}), and is in good agreement with that predicted by a model
treating the pulsar as an oblique rotator with a current-starved outer
magnetosphere (\cite{mel97}).  For the four other pulsars for which it has been
measured, $n$ ranges between 1.4 and 2.8 (see \S~\ref{sec:intro}).
That observed braking indices are smaller than 3 can be explained in a
variety of ways (see \cite{mel97} for a review), including a kinetic
energy-dominated flow at the light cylinder, or an increase in the
magnetic moment of the star over time (\cite{br88}).  None of these
scenarios are consistent with all the observations (\cite{aro92}).
Measurement of $\dddotnu$ would constrain these possibilities further.
At present the upper limit in Table~\ref{tab:parms} is 30 times the
value expected from a simple spin-down law (\cite{br88}).  Whether this
can be measured, and how much the measurement of $n$ can be improved
with further observations, will depend on the level of timing noise and
glitch activity displayed by the pulsar.

Assuming that $P_0 \ll P$, but using the measured values of $P$, $\dot
P$, and $n$ (Table~\ref{tab:parms}) in equation~(\ref{eq:tau}), the age
of PSR~J1119$-$6127 is $1.7\pm0.1$\,kyr, including possible biases due
to timing noise and glitches (see \S~\ref{sec:obs}). Of course, if the
pulsar were born spinning slower, it would be younger.  For $P_0 =
0.2$\,s, half the present period, the age is 1.2\,kyr.  In any case it
is clear that PSR~J1119$-$6127 is among the very youngest neutron stars
known.

Three other pulsars with characteristic ages under 2\,kyr are known:
the Crab pulsar ($\tau_{\rm c} = 1.3$\,kyr), PSR~B1509$-$58 in
G320.4$-$1.2 ($\tau_{\rm c}=1.6$\,kyr), and PSR~B0540$-$69 in the Large
Magellanic Cloud ($\tau_{\rm c}=1.7$\,kyr).  All three are associated
with SNRs.  We have searched for evidence of an SNR near
PSR~J1119$-$6127.  Although none is cataloged (\cite{gre96}), data from
the Molonglo Observatory Synthesis Telescope obtained at a radio
frequency of 843\,MHz (\cite{gcly99}) reveal a faint ring of radius
7$'$ centered on the pulsar.  This could be the expanding blast wave of
the SNR.  Its size would imply an expansion velocity of $\sim
10^4$\,km\,s$^{-1}$, for a distance of 8\,kpc and age of 1.6\,kyr,
reasonable for the blast-wave interpretation if the surrounding medium
is of low density and relatively uniform.  Additional ATCA data show
that the shell has a non-thermal radio spectrum (Crawford et al., in
preparation).  This possible SNR is also X-ray-bright, with its
spectrum described by either a power-law or thermal model, and
additional observations are required to further constrain its
properties (\cite{pkc00b}).  Although the supernova that gave birth to
this pulsar occurred in an era in which celestial events were recorded
by some civilizations, this explosion may have been too far south
and/or too distant or too obscured to have been detected by these
observers.

The glitch observed in PSR~J1119$-$6127 is small compared to most
glitches in most pulsars, with $\Delta \nu/\nu =
(4.4\pm0.4)\times10^{-9}$ (Table~\ref{tab:parms}), but it is of similar
fractional size as three of the five glitches observed in the Crab
pulsar over 23 years (Lyne et al.~2000b\nocite{lsg00}).  It remains to
be seen whether at least some of the change measured in $\dot \nu$ is
permanent, as seen in the Crab glitches.  Unless we were unreasonably
lucky, PSR~J1119$-$6127 glitches more often than the Crab pulsar, but
it is curious that its glitches share some characteristics with those
of the Crab: while its period and magnetic field are approximately 10
times larger than the Crab's, its age, and perhaps therefore its
internal temperature, are similar.

Finally, we compare the PSR~J1119$-$6127 system with some young
pulsar/SNR systems.  For ages $\la 2000$\,yr, the radio luminosity
$L_{\rm R}$ of a synchrotron nebula (``plerion'') with a central pulsar
is a measure of the energy output of the pulsar over its lifetime, due
to the relatively long lifetime of the radiating electrons.  The
plerion X-ray luminosity $L_{\rm X}$, on the other hand, reflects the
current $\dot E$ of the pulsar.  For the Crab and PSR~B0540$-$69,
$L_{\rm X} \sim 0.05 \dot E$, while for PSR~B1509$-$58, $L_{\rm X} \sim
0.01 \dot E$ (see Helfand et al.~1995, and references
therein\nocite{hbw95}).  For SNR 3C58, Helfand et al.\nocite{hbw95}
find that all available data can be reconciled with a (candidate)
pulsar having $P \sim 0.2$\,s, $\dot P \sim 4\times 10^{-12}$
(parameters similar to those of PSR~J1119$-$6127 --- see
Table~\ref{tab:parms}), and with $L_{\rm X} \sim 5 \times 10^{-4} \dot
E$.  For PSR~J1119$-$6127, with a current $\dot E$ 200 times smaller
than the Crab's, the limit on plerionic X-ray emission is $L_{\rm X}
\la 10^{-3} \dot E$ (Pivovaroff et al.~2000a\nocite{pkc00b}).  If
PSR~J1119$-$6127 were born with a small period, it would have had a
much larger $\dot E$ within the past $\sim 1700$\,yr, possibly larger
than the Crab's initially.  That energetic past might be reflected in
plerionic radio emission near the pulsar, depending on the local
environment.  A measurement of $L_{\rm X}$ and $L_{\rm R}$ may in
principle provide information about whether PSR~J1119$-$6127 was born
with a rapid spin rate, as commonly assumed for most pulsars, or
whether it was born a slow rotator.

%Kothes~(1998) \nocite{kot98} finds a very good correlation between
%plerion surface brightness at 1\,GHz, $\Sigma$, and $\dot E/D$, where
%$D$ is the plerion diameter.  For PSR~J1119$-$6127 the predicted
%surface brightness is $\Sigma = 2.7 \times 10^{-20} (8\,\mbox{kpc}/d)
%(1'/\Theta)$\,W\,m$^{-2}$\,Hz$^{-1}$\,sr$^{-1}$, where $d$ is the
%pulsar distance and $\Theta$ is the angular diameter of the plerion.

\subsection{PSR~J1814$-$1744}\label{sec:p1814}

Figure~\ref{fig:ppdot} is a plot of $\dot P$ versus $P$ for the radio
pulsar population.  PSRs~J1119$-$6127 and J1814$-$1744 are indicated,
and we infer $B = 4.1 \times 10^{13}$ and $5.5 \times 10^{13}$\,G
respectively, using equation~(\ref{eq:B}).  These are the highest
magnetic field strengths yet observed among radio pulsars.  The pulsars
with the next largest values of $B$ are PSRs~J1726$-$3530
$(P=1.1\,\mbox{s}; B = 3.7 \times 10^{13}\,\mbox{G})$ and J1632$-$4818
$(P=0.8\,\mbox{s}; B = 2.3 \times 10^{13}\,\mbox{G})$, also discovered
in the multibeam survey\footnote{See
http://www.atnf.csiro.au/$\sim$pulsar/psr/pmsurv/pmwww/pmpsrs.db.}.
Prior to this survey the largest value was $B = 2.1 \times 10^{13}$\,G
for the 2.4\,s PSR~B0154+61 (\cite{antt94}).

Also shown in Figure~\ref{fig:ppdot} are the sources usually identified
as magnetars, namely the five anomalous X-ray pulsars (AXPs) and two
soft gamma repeaters (SGRs) for which $P$ and $\dot{P}$ have been
measured.  AXPs are characterized by X-ray periods in the range
5--12\,s and extremely rapid spin down (\cite{gv98}), while the SGRs
exhibit occasional enormous bursts of $\gamma$-radiation and AXP-like
X-ray pulsations during quiescence.

Most models of the radio emission physics (\cite{mt77}) depend on
pair-production cascades above the magnetic poles and hence on the
magnitude of the magnetic field.  However, at field strengths near or
above the so-called ``quantum critical field,''
\begin{equation}
\label{eq:Bc}
B_{\rm c} \equiv \frac{m_e^2 c^3}{e \hbar} = 4.4 \times 10^{13} \;\; {\rm G},
\end{equation} 
the field at which the cyclotron energy is equal to the electron
rest-mass energy, processes such as photon splitting may inhibit
pair-producing cascades.  It has therefore been argued (\cite{bh98b})
that a radio-loud/radio-quiet boundary can be drawn on the $P$--$\dot
P$ diagram, with radio pulsars on one side, and AXPs and SGRs on the
other (see dotted line in Fig.~\ref{fig:ppdot}).

\medskip
\epsfxsize=8truecm
\epsfbox{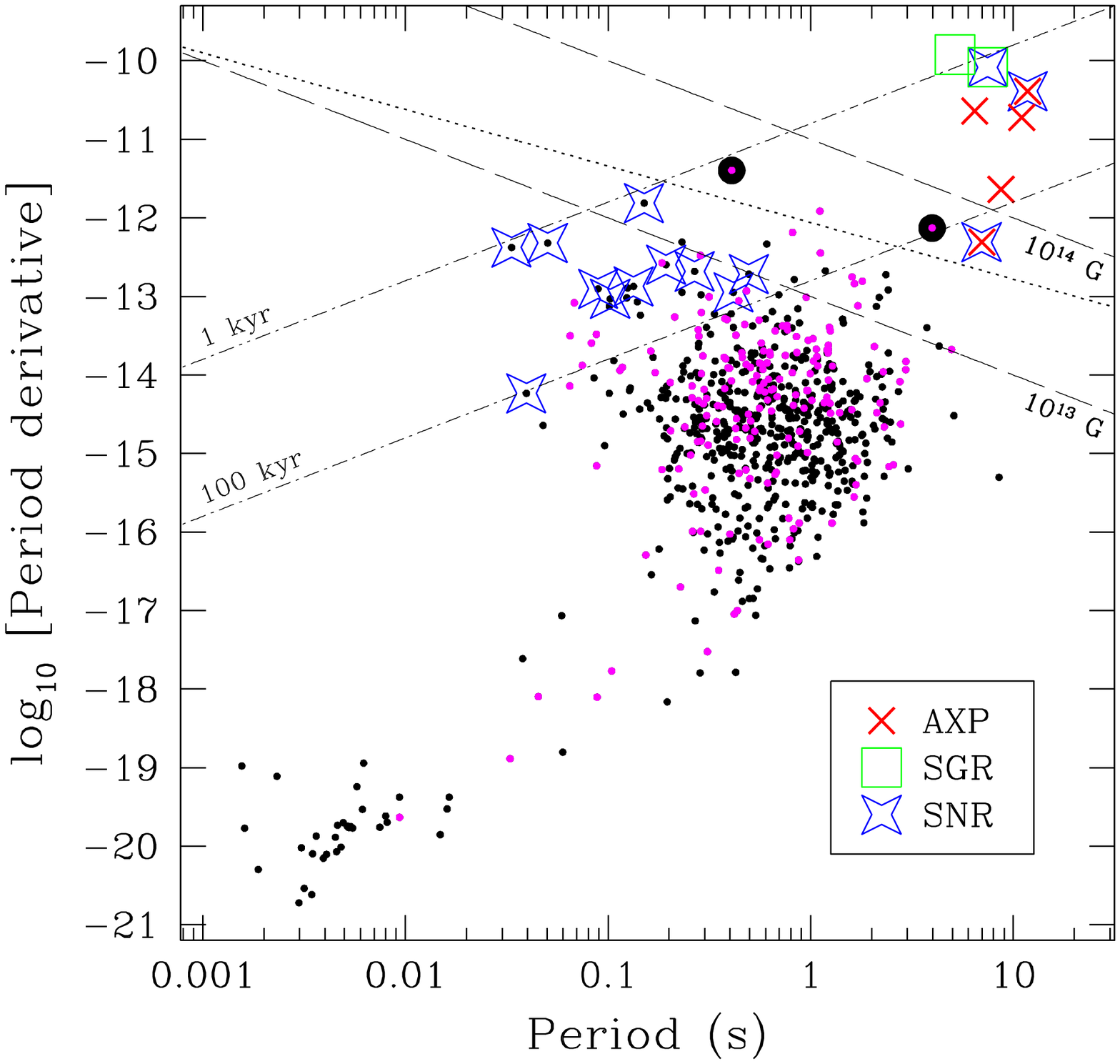}
\figcaption[ppdot.eps]{\label{fig:ppdot}
Plot of $\dot P$ versus $P$ for radio pulsars (dots), anomalous X-ray
pulsars (AXPs), and soft gamma-ray repeaters (SGRs).  PSRs~J1119$-$6127
and J1814$-$1744 are identified by large filled circles, and sources
plausibly associated with supernova remnants (SNRs) are noted.  Lines
of constant characteristic age and surface magnetic field strength are
drawn.  The dotted line shown between the lines for $B=10^{13}$ and
$10^{14}$\,G indicates a hypothesized approximate theoretical boundary
(\cite{bh98b}) separating radio-loud and radio-quiet neutron stars due
to effects relating to magnetic fields close to the critical field
$B_{\rm c}$ (see discussion following equation~[\ref{eq:Bc}]). }
\bigskip

The existence of PSRs~J1119$-$6127, J1726$-$3530, and J1814$-$1744
demonstrates that radio emission can be produced in neutron stars with
$B \ga B_{\rm c}$.  The radio luminosities of these objects (Table~1)
are typical for observed radio pulsars.  Thus, photon splitting does
not appear to inhibit radio emission at these magnetic fields, in
agreement with Usov \& Melrose (1995)\nocite{um95} who argue that this
process is inhibited by polarization selection rules.  Also, there are
both astrophysical and instrumental selection effects which bias
searches against the detection of long-period ($P \ga 5$\,s) radio
pulsars such as J1814$-$1744:  evidence suggests their beams are
narrower (e.g., \cite{ymj99}), so the chances of one intersecting our
line-of-sight are smaller, and instrumental high-pass filtering
intended to remove baseline variations reduces the sensitivity of
searches for such pulsars.  Pulsars such as J1814$-$1744 could
therefore be more prevalent than present numbers suggest.

Especially noteworthy is the proximity of PSR~J1814$-$1744 to the
cluster of AXPs and SGRs at the upper right corner of
Figure~\ref{fig:ppdot}.  In particular, this pulsar has a very similar
$\dot P$ to that of the well-known AXP 1E~2259+586 (\cite{fg81};
\cite{bsss98}), which has a period of 7\,s.  The disparity in their
emission properties is therefore surprising.

The absence of X-ray emission from the direction of PSR~J1814$-$1744,
inferred from archival ASCA and ROSAT observations, implies that it
must be significantly less luminous than 1E~2259+586 (\cite{pkc00}).

The radio emission upper limit for 1E~2259+586 (\cite{cjl94};
\cite{llc98}) implies an upper limit on the radio luminosity at
1400\,MHz of 0.8\,mJy\,kpc$^2$, $10^{-2}$ that of PSR~J1814$-$1744,
assuming a distance of 4\,kpc (\cite{rp97}).  This limit is comparable
to the lowest values observed for the radio pulsar population
(\cite{tnj+94}).  That the radio pulse may be unobservable because of
beaming cannot of course be ruled out.

The radio-loud/radio-quiet boundary line displayed in
Figure~\ref{fig:ppdot} is more illustrative than quantitative
(\cite{bh98b}).  However, the apparently normal radio emission from
PSRs~J1119$-$6127, J1726$-$3530, and J1814$-$1744, and the absence of
radio emission from 1E~2259+586, located very close to PSR~J1814$-$1744
on a $P$--$\dot P$ diagram (Fig.~\ref{fig:ppdot}), suggests that it may
be difficult to delineate any such boundary.

The two sources are also similar in their levels of rotational
stability, at least on time scales of $\sim 2$\,yr: PSR~J1814$-$1744
displays timing noise in the amount expected for a radio pulsar with
its $\dot P$ (see \S~\ref{sec:obs}), as is the upper limit on timing
noise for 1E~2259+586 over a 2--3\,yr span (\cite{kcs99}).  However,
longer term incoherent timing of 1E~2259+586 has revealed significant
deviations from a simple spin-down model.  These have been interpreted
as being evidence for radiative precession of the neutron star, due to
its physical distortion by the strong magnetic field (\cite{mel99}).
Alternatively, Heyl \& Hernquist~(1999)\nocite{hh99} suggest the
deviations are due to extremely large glitches.  In either model,
similar behavior might be expected of PSR~J1814$-$1744; continued radio
timing will be sensitive to it.

The similar spin parameters for these two stars and, in turn, many
common features between 1E~2259+586 and some other AXPs and SGRs,
suggest that very high inferred magnetic field strengths cannot be the
sole factor governing whether or not an isolated neutron star is a
magnetar or a radio pulsar.  Other possible factors include
heavy-element atmospheric composition and youth (\cite{td93a};
\cite{hh97}; see also Pivovaroff et al.~2000b\nocite{pkc00}).  The age
of PSR~J1814$-$1744, if $P_0 \ll P$ and $n=3$, is 85\,kyr.  It is
unlikely that any associated supernova remnant would still be
observable and indeed there is none known in the vicinity
(\cite{gre96}).

We also note that the recently proposed accretion model for AXPs
(\cite{chn00}), in which they are accreting from a fall-back disk
formed from material remaining after the supernova explosion, is
challenged by PSR~J1814$-$1744.  In this model, the neutron star should
not be a radio pulsar, but rather an AXP progenitor in a ``dim
propeller phase,'' its rotational frequency being still too high for
the accreting material to overcome the centrifugal barrier.  Of course,
it is always possible that in this one case no fall-back disk formed.

Proof that AXPs or SGRs are isolated high-magnetic-field neutron stars
would come from either the discovery of magnetar-like emission from a
radio pulsar, or radio pulsations from a putative magnetar.  While such
radio emission was not expected due to theoretical considerations,
because of the high inferred magnetic fields, the discovery of
PSR~J1814$-$1744 shows that this emission does occur at magnetic field
values characteristic of at least some magnetars, opening the
possibility that magnetars also emit observable radio waves.

\acknowledgements

We thank B.~Gaensler for assistance with the MOST data analysis, and
D.~Nice, M.~Pivovaroff, D.~Helfand, L.~Hernquist, and B.~Schaefer for
useful discussions.  The Parkes radio telescope is part of the
Australia Telescope which is funded by the Commonwealth of Australia
for operation as a National Facility managed by CSIRO.  V.~M.~K. and
F.~Crawford are supported by a National Science Foundation CAREER award
(AST-9875897).  F.~Camilo is supported by NASA grant NAG~5-3229.

%\bibliographystyle{apj1d}
%\bibliography{journals_apj,modrefs,psrrefs,crossrefs}

\end{document}